\documentclass[twocolumn,showpacs,preprintnumbers,amsmath,amssymb]{revtex4}

\usepackage{graphicx}
\usepackage{dcolumn}
\usepackage{bm}

\begin{document}


\title{Medium and Small Scale Analysis of Financial Data}

\author{Andreas P. Nawroth}
\author{Joachim Peinke}
\affiliation{%
Institut f\"ur Physik, Carl-von-Ossietzky Universit\"at Oldenburg, D-26111 Oldenburg, Germany
}%

\date{\today}

\begin{abstract}
A stochastic analysis of financial data is presented. In particular we investigate how the statistics of log returns change with different time delays $\tau$. The scale dependent behaviour of financial data can be divided into two regions. The first time-range, the small-timescale region (in the range of seconds) seems to be characterized by universal features. The second time-range, the medium-timescale range from several minutes upwards  and can be characterized by a cascade process, which is given by a stochastic Markov process in the scale $\tau$. A corresponding Fokker-Planck equation can be extracted from given data and provides a non equilibrium thermodynamical description of the complexity of financial data.

\end{abstract}

\maketitle

\section{\label{sec_Intro} Introduction}

One of the remarkable features of the complexity of the financial market is that very often financial quantities display non-Gaussian statistics often denoted as heavy tailed or intermittent statistics, for further details see   \cite{fama1965,mandelbrot1963,clark1973,mantegna1995,castaing1990a,lux1999,bouchaud2001,muzy2000,ghashghaie96}.

\begin{figure}
\includegraphics[width= 8.5cm]{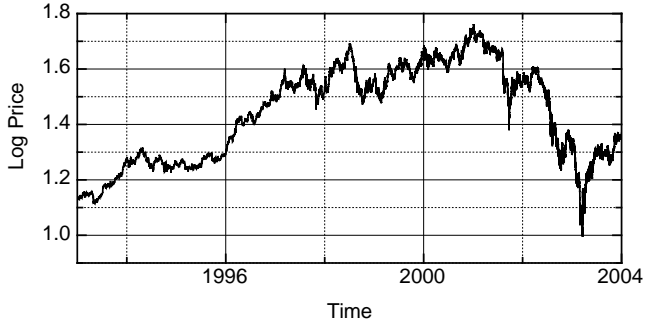}
\caption{\label{fig_ts} Log price for Bayer for the years 1993-2003}
\end{figure}

To characterize the fluctuations of a financial time series $x(t)$, most commonly quantities like returns, log-returns or price increments are used. Here, we consider the statistics of the log return $y(\tau)$ over a certain timescale $\tau$, which is defined as:
\begin{eqnarray}
    y(\tau) \; = \; \log x(t+\tau) - \log x(t). \label{eq_IncDef}
\end{eqnarray}
where $x(t)$ denotes the price of the asset at time $t$. We suppressed the dependence of the log return $y(\tau)$ on the time $t$, since we assume the underlying stochastic process to be stationary. In this paper we present mainly results for Bayer for the time span of 1993 to 2003.  The financial data sets were provided by the Karlsruher Kapitalmarkt Datenbank (KKMDB) \cite{luedecke1998}. The graph of the logarithm of the price time series is shown in Fig. \ref{fig_ts}.

\section{\label{sec_SmallScale} Small scale analysis}

\begin{figure}
\includegraphics[width= 8.5cm]{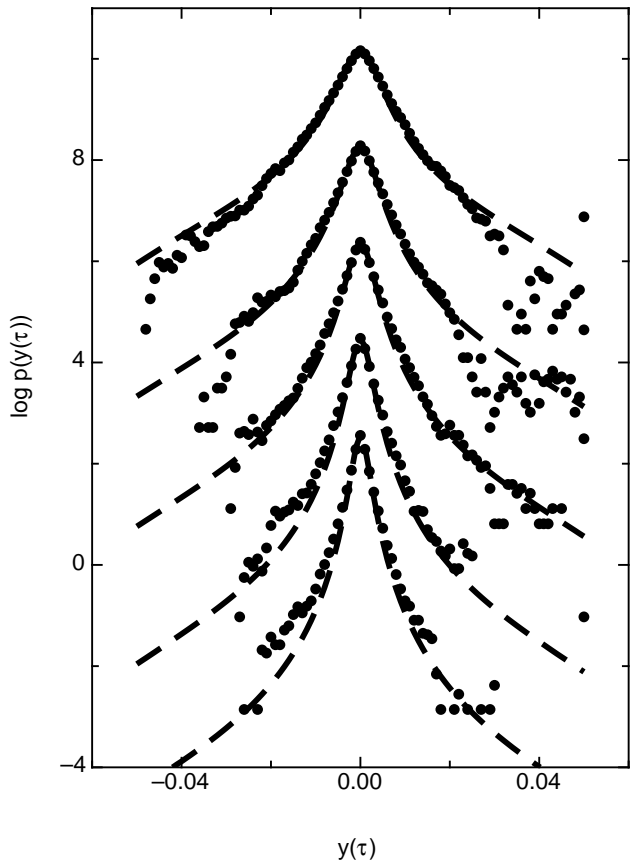}
\caption{\label{fig_ipdfs} Unconditional probability densities $p(y(\tau))$ for the timescales of 
$\tau = 240s, 454s, 955s, 1800s$ and $3766s$ (bottom up) obtained from the original data (dots) and reconstructed from the extracted Fokker-Planck equation (dashed lines).}
\end{figure}

First we look at the statistics of $p(y(\tau))$ as shown in Fig. \ref{fig_ipdfs}. Here we find the remarkable feature of financial data that the probability density functions (pdfs) are not Gaussian, but exhibit heavy tailed shapes. Another remarkable feature is the change of the shape with the size of the scale variable $\tau$. To analyse the changing statistics of the pdfs with the scale $\tau$ a non-parametric approach is chosen. The distance between the pdf $p(y(\tau))$ on a timescale $\tau$ and a pdf $p_T(y(T))$ on a reference timescale $T$ is computed. As a reference timescale, $T = 1 sec$ is chosen. In order to look only at the shape of the pdfs and to exclude effects due to varying mean and variance, all pdfs $p(y(\tau))$ have been normalized to a zero mean and a standard deviation of 1.

\begin{figure*}
\includegraphics[width= 16cm]{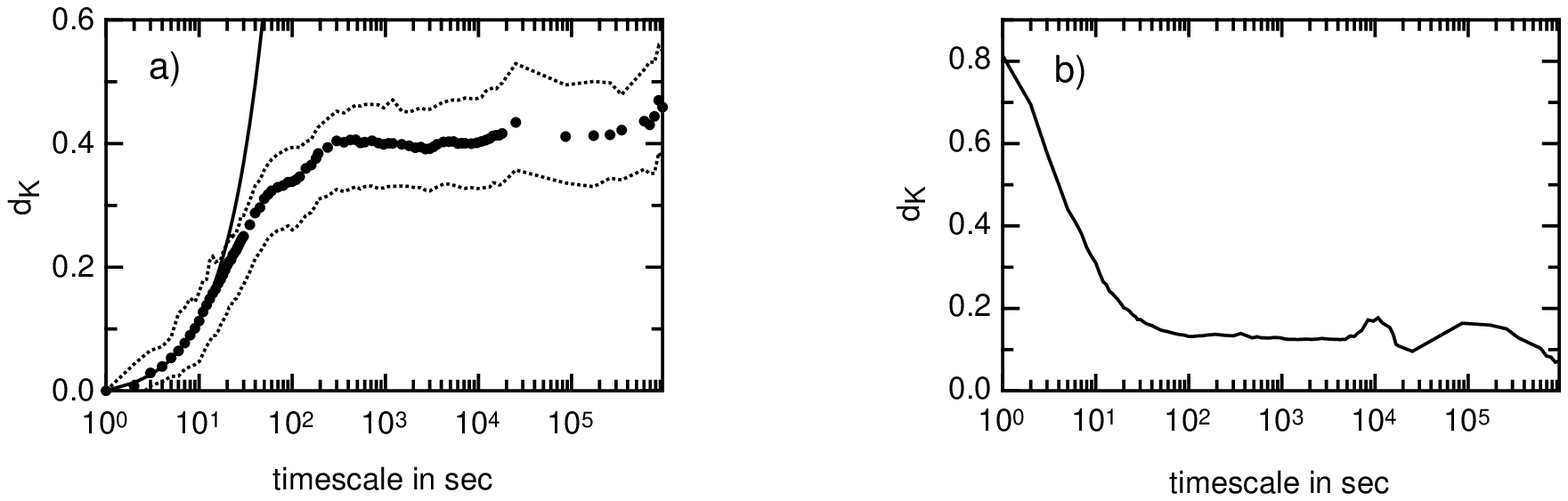}
\caption{\label{fig_dK} Distance measure $d_K$ for a reference distribution $p_{T}(y)$ for Bayer. a) As reference timescale $T=1 sec$ is chosen. The bold dots represent the estimated value, the dotted lines the one-sigma error bound and the solid line the linear fit for the first region, after \cite{nawroth2006}. b) As a reference distribution $p_{T}(y)$ a normalised Gaussian distribution is chosen.}
\end{figure*}

As a measure to quantify the distance between two distributions $p(y(\tau))$ and $p_{T}(y(T))$, the Kullback-Leibler-Entropy  \cite{kullback1968}
\begin{eqnarray}
d_K(\tau) := \int \limits^{+\infty}_{-\infty}dy \; p(y(\tau)) \cdot \ln \left ( \frac{p(y(\tau))}{p_{T}(y(T))} \right )
\label{eq_distance_KL}
\end{eqnarray}
is used. In Fig. \ref{fig_dK} the evolution of $d_K$ with increasing $\tau$ is shown, which measures the change of the shape of the pdfs. For different stocks we found that for timescales smaller than about one minute a linear growth of the distance measure seems to be universally present, see Fig. \ref{fig_dK}a. If as a reference distribution a normalised Gaussian distribution is taken, the fast deviation from the Gaussian shape in the small timescale regime becomes evident, as displayed in Fig. \ref{fig_dK}b. The independence of this small scale behaviour on the particular choice of the measure and on the choice of the stock is shown in \cite{nawroth2006}.

\section{\label{sec_MediumScale} Medium scale analysis}

Next the behaviour for larger timescales ($\tau > 1 min$) is discussed. Here we proceed the analysis with the idea of a cascade. As has been shown by \cite{ghashghaie96,friedrich00b,renner01b} it is possible to grasp the complexity of financial data by cascade processes running in the variable $\tau$. In particular it has been shown that it is possible to estimate directly from given data a stochastic cascade process in form of a Fokker-Planck equation \cite{friedrich00b,renner01b}. The underlying idea of this approach is to access statistics of all orders of the financial data by the general joint n-scale probability densities
$ p(y_{1},\tau_{1};y_{2},\tau_{2}; ... ;y_{N},\tau_{N}) $ (Here we use the shorthand notation $y_1=y(\tau_1)$ and take without loss of generality $\tau_i<\tau_{i+1}$. The smaller log returns $y(\tau_{i})$ are nested inside the larger log returns $y(\tau_{i+1})$ with common end point $t$.)

The joint pdfs can be expressed as well by the multiple conditional probability densities $p(y_{i},\tau_{i}|y_{i+1},\tau_{i+1}; ... ;y_{N},\tau_{N})$. This very general n-scale characterization of a data set, which contains the general n-point statistics, can be simplified essentially if there is a stochastic process in $\tau$, which is a Markov process. This is the case if the conditional probability densities fulfil the following relations:
\begin{eqnarray}
	p(y_{1},\tau_{1}|y_{2},\tau_{2};y_{3},\tau_{3};\ldots
	y_{N},\tau_{N})  & = & 
	p(y_{1},\tau_{1}|y_{2},\tau_{2}). \;  \label{markovcondtheo}
\end{eqnarray}

Consequently,
\begin{eqnarray}\label{eq_chain}
\lefteqn{p(y_1,\tau_1;...;y_N,\tau_N) =  } \hspace{-0.0cm} \\
\nonumber \\
& & p(y_1,\tau_1|y_2,\tau_2)\cdot ...\cdot p(y_{N-1},\tau_{N-1}|y_N,\tau_N)\cdot p(y_N,\tau_N) \nonumber 
\end{eqnarray}
holds.

Equation (\ref{eq_chain}) indicates the importance of the conditional pdf for Markov processes.  Knowledge of $p(y,\tau|y_{0},\tau_{0})$ (for arbitrary scales $\tau$ and $\tau_{0}$ with $\tau<\tau_{0}$) is sufficient to generate the entire statistics of the increment, encoded in the N-point probability density $p(y_{1},\tau_{1};y_{2},\tau_{2}; \ldots ;y_{N},\tau_{N}) $.

For Markov processes the conditional probability density satisfies a master equation, which can be put into the form of a Kramers-Moyal expansion for which the Kramers-Moyal coefficients $D^{(k)}(y,\tau)$ are defined as the limit $\Delta \tau \rightarrow 0$ of the conditional moments $M^{(k)}(y,\tau,\Delta \tau)$:
\begin{eqnarray}\label{eq_DnDef}
	D^{(k)}(y,\tau) = \lim_{\Delta \tau \rightarrow 0} \, M^{(k)}(y,\tau,\Delta\tau)
\end{eqnarray}
\begin{eqnarray}\label{eq_MnDef}
\lefteqn{M^{(k)}(y,\tau,\Delta \tau) =  } \hspace{-0.0cm} \\
\nonumber \\
& & \frac{\tau}{k! \, \Delta \tau} \, \int\limits_{-\infty}^{+\infty} \, \left( \tilde{y} - y \right)^k \,
	p\left( \tilde{y}, \tau - \Delta \tau | y,\tau \right) \, d \tilde{y} . \nonumber 
\end{eqnarray}

For a general stochastic process, all Kramers-Moyal coefficients are different from zero. According to Pawula's theorem, however, the Kramers-Moyal expansion stops after the second term, provided that the fourth order coefficient $D^{(4)}(y,\tau)$ vanishes. In that case, the Kramers-Moyal expansion reduces to a Fokker-Planck equation (also known as the backwards or second Kolmogorov equation):
\begin{eqnarray}\label{eq_FokPla}
\lefteqn{- \tau \frac{\partial}{\partial \tau} \, p(y,\tau|y_{0},\tau_{0}) =  } \hspace{-0.0cm} \\
\nonumber \\
& & \left\{ \, - \frac{\partial}{\partial y} D^{(1)}(y,\tau) \, + \,
        \frac{\partial^2}{\partial y^2} D^{(2)}(y,\tau) \, \right\}
        p(y,\tau|y_{0},\tau_{0}) . \nonumber 
\end{eqnarray}
$D^{(1)}$ is denoted as drift term, $D^{(2)}$ as diffusion term. The probability density $p(y,\tau)$ has to satisfy the same equation, as can be shown by a simple integration of Eq. (\ref{eq_FokPla}).

\section{\label{sec_RFB}Results for Bayer}

\begin{figure}[h!]
\includegraphics[width= 8.5cm]{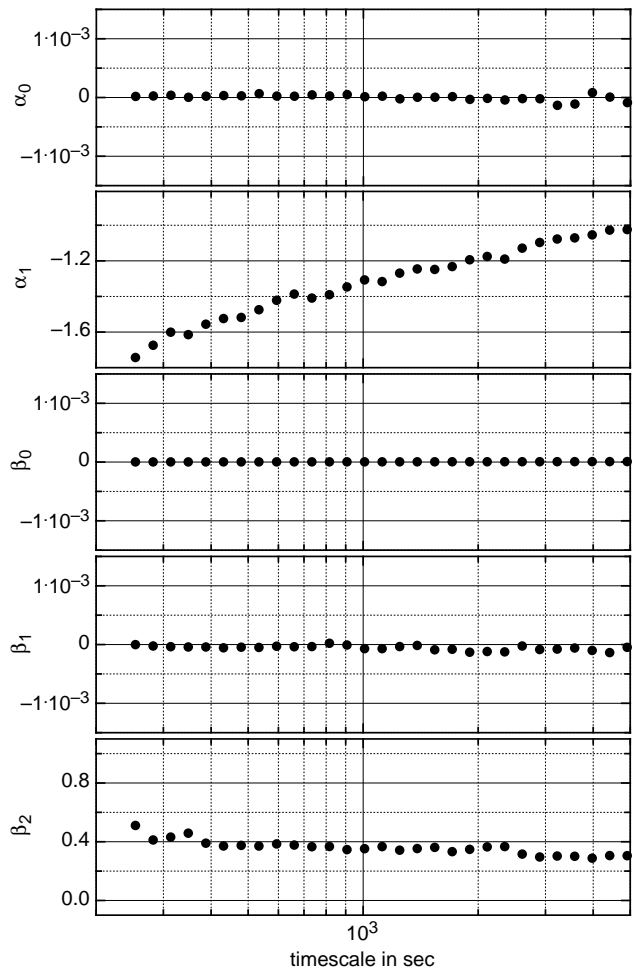}
\caption{\label{fig_000Coeff} The parameters $\alpha_0,\alpha_1,\beta_0,\beta_1$ and $\beta_2$ of the parameterisation of the Kramers-Moyal coefficients used for the reconstruction.}
\end{figure}

From the data shown in Fig. \ref{fig_ts} the Kramers-Moyal coefficients were calculated according to Eqs. (\ref{eq_MnDef}) and (\ref{eq_DnDef}). Hereby we divided the timescale into intervals
\begin{eqnarray}
\left[\frac{1}{2}(\tau_{i-1}+\tau_{i}),\frac{1}{2}(\tau_{i}+\tau_{i+1})\right[ \nonumber
\end{eqnarray}
assuming that the Kramers-Moyal coefficients are constant with respect to the timescale $\tau$ in each of these sub intervals of the timescale. We started with a smallest timescale of $240s$ and continued in such a way that $\tau_{i}=0.9\cdot \tau_{i+1}$. The Kramers-Moyal coefficients themselves were parameterised in the following form:
\begin{eqnarray}
\label{eq_markov_properites}
D^{(1)} & = & \alpha_0 + \alpha_1 y\\
D^{(2)} & = & \beta_0 + \beta_1 y + \beta_2 y^2.
\end{eqnarray}
The coefficients we obtained by this procedure are shown in Fig. \ref{fig_000Coeff}. This result shows that the rich and complex structure of financial data, expressed by multiscale statistics, can be pinned down to coefficients with a quite simple functional form.

\begin{figure}
\includegraphics[width= 8.5cm]{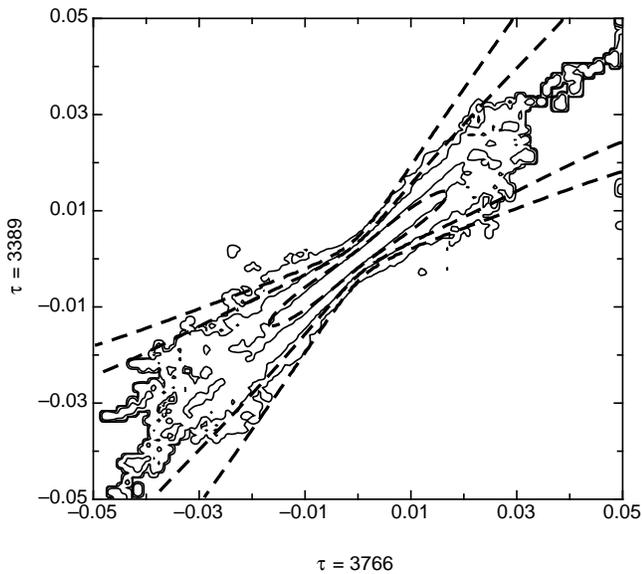}
\caption{\label{fig_000CondPdf_3766_3389} Conditional probability density $p(y(\tau = 3389s)|y(\tau = 3766s))$ of given data (unbroken lines) and reconstructed by the numerical solution of the Fokker-Planck equation (broken lines). }
\end{figure}

To show the quality of our results we reconstruct the measured statistics by the estimated Fokker-Planck equations.
At first, the conditional probability densities $p(y(\tau_i)|y(\tau_{i+1}))$ were reconstructed. As an example the conditional probability density $p(y(\tau = 3389s)|y(\tau = 3766s))$ is shown in Fig. \ref{fig_000CondPdf_3766_3389}. The reconstructed conditional probability density and the one calculated directly from the data are in good agreement.
As a next step we used the pdf on the scale of $\tau=27900s$ and the reconstructed conditional probability densities to calculate the increment pdfs on timescales between four minutes and one hour. The results for the timescales of $\tau = 3766s, 1800s, 955s, 454s$ and $240s$ are shown in Fig. \ref{fig_ipdfs}. Again the agreement between unconditional probability densities $p(y(\tau))$ of the original data (dots) and the reconstructed ones (broken lines) is very good.

\section{\label{sec_Discussion} Discussion}

The results indicate that for financial data there are two scale regimes. In the small scale regime the shape of the pdfs change very fast and a measure like the Kullback-Leibler entropy increases linearly. At timescales of a few seconds not all available information may be included in the price and processes necessary for price formation take place. Nevertheless this regime seems to exhibit a well defined structure, expressed by the very simple functional form of the Kullback-Leibler entropy with respect to the timescale $\tau$.

Based on a stochastic analysis we have shown that a second time range, the medium scale range exists, where multiscale joint probability densities can be expressed by a stochastic cascade process. Here the information on the comprehensive multiscale statistics can be expressed by simple conditioned probability densities. This simplification may be seen in analogy to the thermodynamical description of a gas by means of statistical mechanics. The comprehensive statistical quantity for the gas is the joint n-particle probability density, describing the location and the momentum of all the individual particles. One essential simplification for the kinetic gas theory is the single particle approximation. The Boltzmann equation is an equation for the time evolution of the probability density $p(\mathbf{p}, t)$ in one-particle phase space, where x and $\mathbf{p}$ are position and momentum, respectively. In analogy to this we have obtained for the financial data a Fokker-Planck equation for the scale $\tau$ evolution of conditional probabilities, $p(y_{i},\tau_{i}|y_{i+1},\tau_{i+1})$. In our cascade picture the conditional probabilities can not be reduced further to single probability densities, $p(y_{i},\tau_{i})$, without loss of information, as it is done for the kinetic gas theory.

As a last point we want to mention that based on the information of the Fokker-Planck equation it is possible to generate artificial data sets. As pointed out in \cite{nawroth2006a}, the knowledge of conditional probabilities can be used to generate time series. One important point is that one uses increments $y(\tau)$ with common right endpoints. By the knowledge of the n-scale conditional probability density of all $y(\tau_i)$ the stochastically correct next point can be selected. We could show that time series for turbulent data generated by this procedure even reproduces quite well the conditional probability densities, as the central quantity for a comprehensive multiscale characterization.

\begin{acknowledgments}
 
For helpful discussion we want to thank R. Friedrich, Ch. Renner, D. Sornette.

\end{acknowledgments}



\bibliography{AG_LITERATUR}

\end{document}